# Anisotropy of Laser-Induced Bulk Damage of Single Crystals

Krupych O., Dyachok Ya., Smaga I., Vlokh R.

Institute of Physical Optics, 23 Dragomanov St., 79005 Lviv, Ukraine
e-mail: vlokh@ifo.lviv.ua

## Abstract

The regularities of laser-induced damage of anisotropic materials, such as $LiNbO_3$ and KDP dielectric single crystals, are experimentally studied. It is revealed that the shape of laser-induced damage in the dielectric crystals depends on the elastic symmetry of crystal and the propagation direction of the laser beam. When the beam propagates along the optic axis of crystals, the figures of the laser damage are six-path stars for $LiNbO_3$ and four-path ones for KDP crystals. For the direction parallel to X and Y axes in KDP crystal, the damage has initially cross-like configuration, with further splitting of Z-oriented crack into two cracks in the process of damage evolution, leading to transformation of orthogonal-type damage to a hexagonal-type one.

## Introduction

Studies of bulk laser-induced damage of optical materials have been usually directed at measuring the damage threshold [1-4] and analysing the damage mechanism [5-7]. Only few researchers have studied bulk damage morphology, when irradiating the crystals with high-power laser pulses [8-10]. It has been marked that, in case of propagation along the optic axis, the laser-induced damages have a star-like view, if observed perpendicular to the optic axis. These stars would generally correspond to the order of the corresponding symmetry axis. When the laser beam propagates perpendicular to the higher-order symmetry axis, the literature data become ambiguous. For example, H. Yoshida et al. [10] have revealed that the laser beam directed along the X axis in potassium dihydrogen phosphate (KDP, $KH_2PO_4$) crystals yields in eight-path stars. However, Yoshimura M. et al. [9] have observed six-path damage stars in CLBO crystals for the identical experimental geometry, though these crystals have exactly the same symmetry ($\bar{4}2m$) as KDP. Thus, the goal of this work is to study regularities of the laser-induced damage in anisotropic materials on the example of dielectric crystals.

## Experimental results and discussion

For studying the optical damage, we used a setup described earlier in [4] and applied a pulsed $Nd^{3+}$ laser ($\tau = 6$ ns; the output radiation energy 60 mJ). The laser beam was focused to the spot with the diameter of about 108 μm, corresponding to the $1/e^2$ intensity level. The focus plane was positioned at the depth of 3–5 mm from the sample face. For each laser beam direction, we performed several shots and shifted the specimen between the shots in order to avoid a damage accumulation.

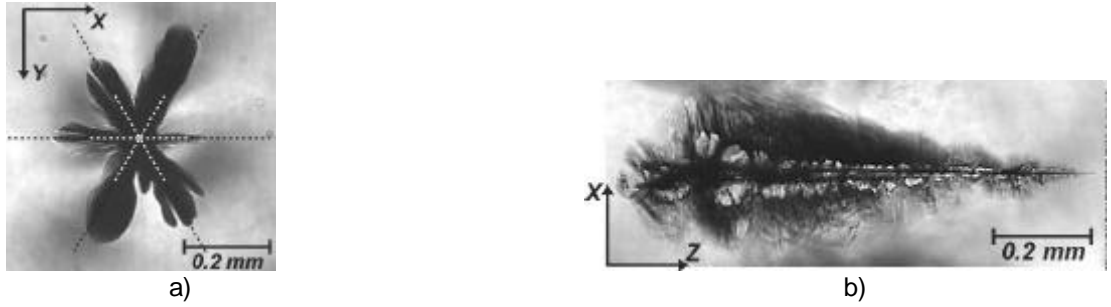

**Fig. 1.** View of the damage tracks for Z-cut LN crystals (a) and the same damage track observed from Y plane (b). The light propagates along Z axis.

To study a bulk damage anisotropy, we used lithium niobate (LN, $LiNbO_3$) and KDP crystals.

The specimen of LN was cut out perpendicular to the principal directions [100], [010] and [001]. It represented a parallelepiped with the dimensions of $8(a) \times 10(b) \times 8.3(c)$ mm$^3$. The light propagated along the Z axis. The resulting damages are shown in Fig. 1. The obtained damage stars viewed along the optic axis manifest nearly hexagonal shape (see Fig. 1a). Such the configuration of cracks corresponds to the point symmetry 3m of LN crystals. The damage channel has an arrowhead shape, with a wide part located near the light focus point and a thinning oriented depthward the sample (Fig. 1b).

The specimen of KDP crystals had nearly cubic shape, with the dimensions $11.2(a) \times 11.2(b) \times 10.8(c)$ mm$^3$ and the faces (100), (010) and (001). We performed the damage experiments using the light directed along all of the three principal axes. If the laser beam propagates along the optic axis (Z direction), the damage stars have the shape of precise rectangular cross, with the rays directed parallel to X and Y axes (Fig. 2a). Our results differ from those reported by H. Yoshida et al. [10], who have observed cross-type stars rotated by 45° around the Z axis, with the crack planes (110) and ($1\bar{1}0$). The difference might be explained by different choices of crystallographic frames of reference. In our case, X and Y axes are oriented along the two-fold symmetry axes. However, sometimes one finds in the literature that the X and Y axes in KDP are directed perpendicular to the mirror planes, which are rotated by 45° with respect to the two-fold axes. Nevertheless, the damage stars correspond to tetragonal group $\bar{4}2m$ in both cases.

H. Yoshida et al. [10] have observed eight-path stars from $a$ plane for the laser beam direction [100]. The crack projections on $a$ plane observed in this experiment subtend the angles of 20°, 47° and 66°. This fact seemingly agrees with neither the orientation of symmetry elements nor the experimental geometry. However, we have noticed an interesting fact. If the authors mentioned above used the alternative crystallographic reference frame and the elementary cell parameters [11] $a=b=10.543$ Å, $c$=6.959 Å, then it would be easy to calculate that the angle between the cell diagonals is equal to 66.8°. This correlates well with the angle between the crack projections reported by H. Yoshida et al. When the laser beam in our experiments is directed along the X axis ([100] direction), we obtain the damage stars close to regular six-point ones (see Fig.3). The orientation of the axes, as well as the reference hexagonal star (dashed lines), is shown in Fig. 3a. One can see that one crack plane is perpendicular to Z axis and the two others are symmetric with respect to this axis (as well as the (010) plane which contains Z axis). However, we have noticed in this case that the damage has initially cross-type view (Fig 3a), with the rays parallel to Y and Z axes. The damage evolution depthward the sample leads to splitting of Z-oriented crack into two cracks (Fig. 3, a to d). In this manner, the initially orthogonal-type damage transforms to a hexagonal-type one. Quite similar damage behaviour is observed for the case of laser beam directed along the Y axis (see Fig. 4). In this case, the initially Z-oriented crack splits into two and forms a hexagonal star, too.

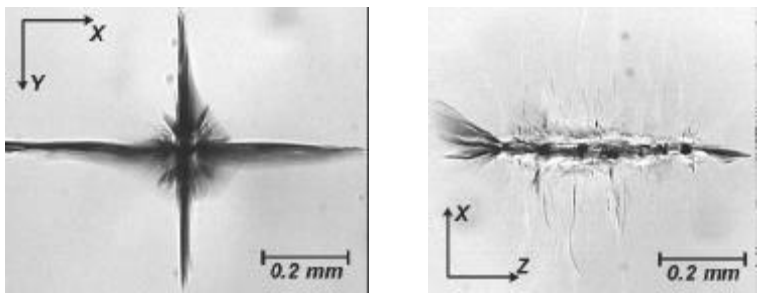

**Fig. 2.** View of the damage tracks for Z-cut KDP crystal (a) and the same damage track observed from Y plane (b). The light propagates along Z axis.

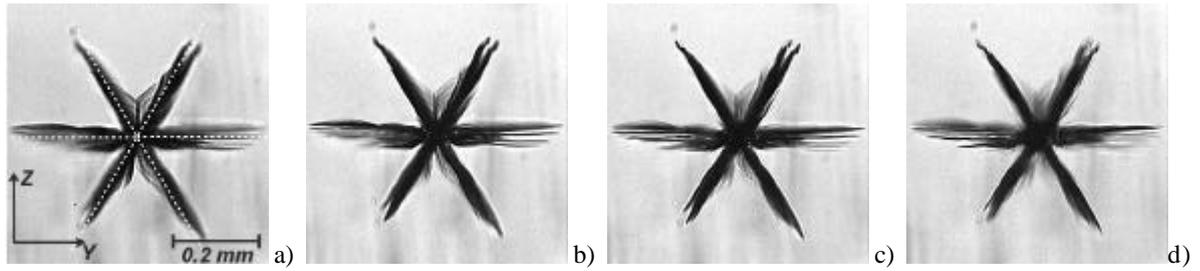

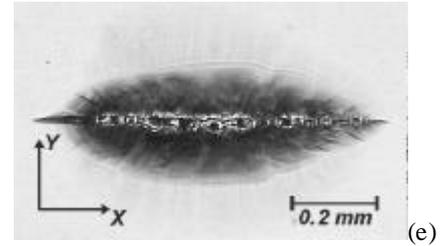

**Fig. 3.** View of the damage tracks for X-cut KDP crystal (a) – (d) and the same track observed from Z plane (e). The focus plane corresponds to (a) exactly the damage origin and (b) 0.15, (c) 0.30 and (d) 0.45 mm behind the origin. The light propagates along X axis.

The damage channels for the KDP crystals observed along perpendiculars to the laser beam directions are shown in Fig. 2b, 3e and 4e. Unlike the LN crystals, the damage traces in the KDP crystals have almost symmetric shape along the laser beam path. Moreover, the damage channels in KDP possess noticeable periodically situated nodes, indicating at a self-focusing character of damages.

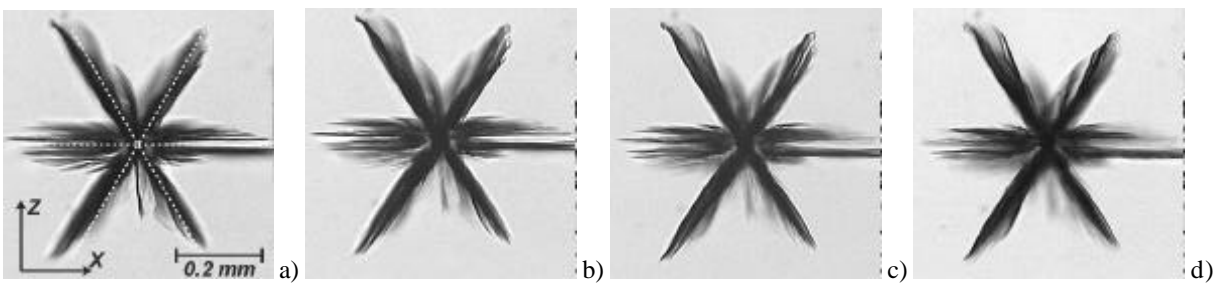

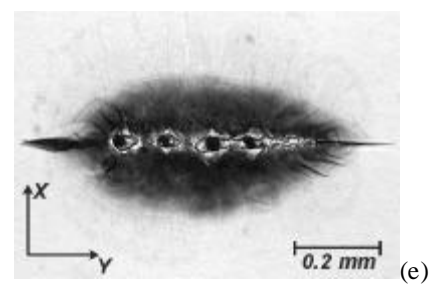

**Fig. 4.** View of the damage tracks for Y-cut KDP crystal (a) – (d) and the same track observed from Z plane (e). The focus plane corresponds to (a) exactly the damage origin and (b) 0.15, (c) 0.30 and (d) 0.45 mm behind the origin. The light propagates along Y axis.

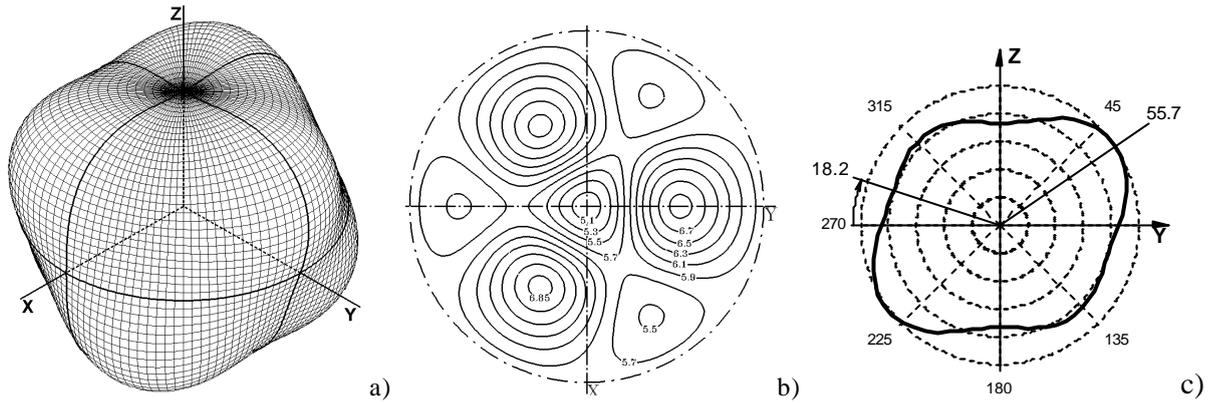

**Fig. 5.** Symmetry of the "extension factor" $E^{-1}$ for the LN crystals: the indicative surface (a), its stereographic projection (b) and X-cut (c).

To explain the observed shape of damage, we have built the indicative surfaces of the "extension factor" $E^{-1}$ (with E being the Young modulus) and its stereographic projections for the studied crystals (see Fig. 5 and 6), using the coefficients of elastic compliances taken from [12]. From the experiment, we obtain that the damage star in the LN crystals, for the laser light direction [001], has the rays parallel to the X axis (Fig. 1a). However, the extension factor in this direction does not acquire extreme values. It would be suggested that the damage cracks are perpendicular to the directions of extreme values of the extension factor (i.e., Y axis in our case). The analysis of the damage stars and the elastic symmetry for KDP allows one to make clear which extrema of the extension factor might be connected with the cracks directions. As seen from Fig. 6b, the extension factor has its minima in the directions of X and Y axes and the maxima in the directions [110] and [1$\bar{1}$0]. The experiment (see Fig. 2a) gives us the orientations of the crack planes [100] and [010], when the laser beam has the direction [001]. This means that the cracks are perpendicular to the directions of *minimal* values of the extension factor. The conclusion could be confirmed by the following fact. As seen from the X-cut of indicative surface of the extension factor for LN crystals (Fig. 5c), the direction of minimal value is inclined by 18.2° to XOY plane. Thus, the cracks obtained with the laser beam direction [001] must be oblique to the Z axis. It leads to wide star paths observed experimentally on the Z-cut LN crystals (Fig. 1a). Unlike the case of NL, the directions of minimal values of the extension factor in KDP are perpendicular to the Z axis (Fig. 6 and 6c). Consequently, the cracks really observed on the Z-cut KDP crystals are sharp (see Fig. 2a).

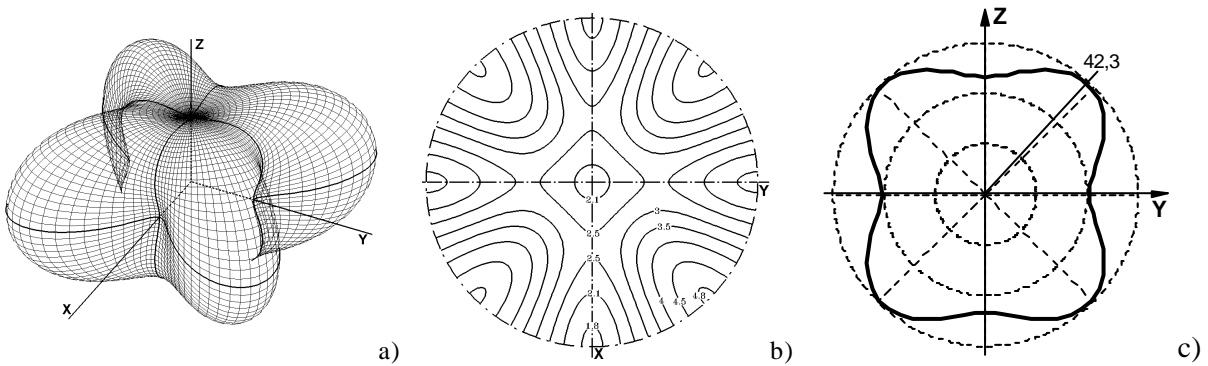

**Fig. 6.** Symmetry of the "extension factor" $E^{-1}$ for the KDP crystals: the indicative surface (a), its stereographic projection (b) and X-cut (c).

Therefore, the results mentioned above enable us to conclude that the shape of damage for the laser beam directed along the optic axis in uniaxial crystals is explained by the elastic symmetry of crystals and corresponds to the order of the symmetry axis. As some directions in crystals are more "compliant" and the other stiffer, the crack orientations are not random. As a result of symmetry, such the directions repeat N times per revolution, where N is the symmetry axis order (see Fig. 5b and 6b). Thus, the laser-induced damage cracks form symmetric stars defined eventually by the elastic symmetry of crystal.

When the laser beam propagates perpendicular to the optic axis in an optically uniaxial crystal, the figures of damage cracks are more complicated. First, the damage tracks depend on the direction of laser beam ([100] or [110]). Second, the symmetry of the damage pattern does not correspond to the symmetry of crystal in the given direction. Although the damage begins with appearance of two perpendicular cracks, its development leads to splitting one of the cracks (which is parallel to the Z axis) into two cracks and forming a hexagonal star.

## Conclusion

Our experimental results indicate that the shape of laser-induced damage of the dielectric single crystals depends on the elastic symmetry of crystal and the laser beam direction. When the laser beam propagates along the optic axis in optically uniaxial crystals, the figures of the laser damage are star-like, with the number of rays defined by the order of symmetry axis.

## Acknowledgement

The authors are grateful to the Ministry of Education and Science of Ukraine (the Project N0104U000460) for financial support of this study.